\documentclass{article}

\usepackage[sorting=none,sortcites=true,minbibnames=3,maxbibnames=3,autopunct=true,hyperref=true,abbreviate=false,backref=true,backend=biber]{biblatex}
\usepackage[final,nonatbib]{neurips_2023}

\usepackage{tikz}
\usepackage[utf8]{inputenc} % allow utf-8 input
\usepackage[T1]{fontenc}    % use 8-bit T1 fonts
\usepackage{hyperref}       % hyperlinks
\usepackage{url}            % simple URL typesetting
\usepackage{booktabs}       % professional-quality tables
\usepackage{amsfonts}       % blackboard math symbols
\usepackage{nicefrac}       % compact symbols for 1/2, etc.
\usepackage{microtype}      % microtypography
\usepackage{xcolor}         % colors
\usepackage{multirow}

\usepackage{subcaption}

\newcommand{\ignore}[1]{}

\newcommand*\circled[1]{\tikz[baseline=(char.base)]{
            \node[shape=circle,draw,inner sep=1pt] (char) {#1};}}
\title{LLeMpower: Understanding Disparities in the Control and Access of Large Language Models}

\author{%
  Vishwas Sathish*, Hannah Lin*, Aditya K Kamath*, Anish Nyayachavadi\\
  \{vsathish, hannahyl, akkamath, anishnya\}@cs.washington.edu
}

\begin{document}

\maketitle

\begin{abstract}
Large Language Models (LLMs) are a powerful technology that augment human skill to create new opportunities, akin to the development of steam engines and the internet. 
% These capabilities expand and improve as the models grow. 
% Access to LLMs can provide a significant boost in productivity to individuals and organizations. 
However, LLMs come with a high cost. 
They require significant computing resources and energy to train and serve. Inequity in their control and access has led to concentration of ownership and power to a small collection of corporations. 
% Unfortunately, these costs scale with model size monopolizing their control to a small group. 
In our study, we collect training and inference requirements for various LLMs. 
We then analyze the economic strengths of nations and organizations in the context of developing and serving these models.
Additionally, we also look at whether individuals around the world can access and use this emerging technology. 
We compare and contrast these groups to show that these technologies are monopolized by a surprisingly few entities. 
We conclude with a qualitative study on the ethical implications of our findings and discuss future directions towards equity in LLM access.
\end{abstract}

\section{Introduction} \label{sec:intro}
LLMs have become ubiquitous in AI research and are gradually permeating our daily lives. 
We have seen LLMs impact developments in many different fields, including journalism~\cite{ding2023harnessing}, medicine~\cite{thir2023medicine}, biology~\cite{brahmavar2023generating}, chemistry~\cite{jablonka202314}, law~\cite{cui2023chatlaw}, education~\cite{moore2023empowering, xiao2023evaluating, tian2024debugbench}, software engineering~\cite{hou2023large, bairi2023codeplan} and robotics~\cite{shu2024llms, zeng2023large}. 
Their widespread adoption promises faster delivery of consumer products and services.
Analyzing the potential for generative AI to transform a wide range of industries, Mckinsey \& Company predict that LLMs will impact productivity and the global economy by trillions of dollars~\cite{chui2023economic}. However, will this economic growth be experienced by all or only a few? We approach this question by examining the economic accessibility of LLMs to various groups of people around the world and by discussing the ethical implications of our findings.

With the emergence of the Transformer architecture~\cite{Vaswani:Neurips:2017} and its subsequent language model BERT~\cite{Devlin:arxiv:2019}, LLMs have been increasing exponentially in size~\cite{minaee2024large}. Each larger model has led to new model capabilities, and recent scaling laws have confirmed that increasing model size leads to more powerful LLMs~\cite{kaplan2020scaling, hoffmann2022training}. However, this rapid growth comes at the cost of increasing energy and hardware requirements. 
When models grow larger, more compute and memory is needed to store model parameters and compute weight updates.
Moreover, LLMs require specialized hardware such as GPUs and TPUs, which come with exorbitant price tags.
% even when accessed through cloud providers. 
As shown in Table \ref{tab:llm_costs}, training GPT-4 took an estimated \$100 million. 
Serving GPT-4 for inference requires roughly \$900 per day. The high compute costs of LLMs create a barrier to access for those with lower financial resources. 

Due to this financial barrier, future breakthroughs in training and serving LLMs are restricted to a few small groups. 
For example, although journalism has been shown to benefit from LLMs~\cite{petridis2023anglekindling}, newsrooms in Africa lack financial resources to train and use dedicated LLMs~\cite{munoriyarwa2023artificial}. 
Similarly, while some have been able to fine-tune their own LLM for use in medical settings~\cite{chen2023meditron70b}, many hospitals do not have the adequate funding required to fine-tune such models for large-scale use. 
Observing these disparities, we analyze the economic resources of several groups around the world, and compare them to the costs of developing and operating LLMs. 

We choose to focus our work on a few key stakeholders:
\emph{Private sector organizations and industries} are the largest group of LLM developers that will ultimately affect the global economy. 
Whether they effectively incorporate and sell LLM-based services will depend on advancements in \emph{research institutions}. 
As products based off LLMs are released, \emph{individuals} will start to form a significant portion of their user base. As such, we look at the distribution of economic resources across (1) small and large corporations, (2) research institutions, and (3) individuals by country.

\begin{figure}
    \centering
    \includegraphics[width=\linewidth]{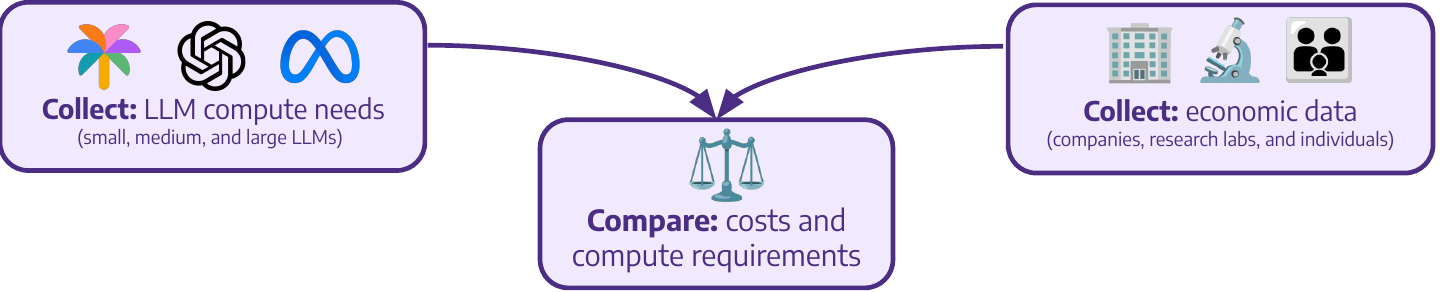}    
    \caption{We compare the costs of 6 LLM models ranging from small to large with the economic resources of companies, research labs, and individuals. (Logos from~\cite{openaiBrand, wikiPalm, metaBrand}.)}
    \label{fig:methods}
\end{figure}

Our work uses the GPU compute cost for these models as a proxy for their practical deployment. 
We do not consider the costs for hiring engineers, data scientists and miscellaneous overhead costs that may be necessary in training or using an LLM. 
We start with this assumption from the observation that the current costs for training LLMs far dwarf other operational costs.
Our contributions can be summarized as follows:
\begin{itemize}
    \item Tabulate the training and inference costs for a few common LLMs. Collect the raw data for financial resources of a range of companies, research institutions, and individuals, from reputed resources available on the internet.
    \item Compare the LLM training costs to the financial resources of companies and research institutions to determine which groups have the ability to dictate the future of this technology.
    \item Analyze how accessible LLMs are to individuals of different countries based on economic and language barriers.
    \item Provide a discussion on ideas for increasing accessibility to LLMs globally.
\end{itemize}

We begin by discussing the related work on LLM equity in Section \ref{sec:related_work}. 
We then describe the methodology used for selecting models and populations to analyze (Section \ref{sec:methods}). 
The results of our analysis are presented in Section \ref{sec:results}, and we follow with a discussion of the implications of our findings. We then briefly mention our suggestions for next steps to diffuse the concentration of power and bridge LLM equity across groups (Section \ref{sec:discussion}).

% \section{Implications of LLM Growth}
\section{Related Work}\label{sec:related_work}

In this section, we briefly discuss prior research work around the wider topic of LLMs and their socio-economic impact. Following the discussion in the introduction, the widespread adoption of LLMs promise faster delivery of consumer products and services powered by AI. This rapid growth comes at the cost of increased energy consumption, hardware resource requirements and human labor.

\textbf{Public Availability:} While the tech industry mega-corporations have traditionally held a monopoly over these large-scale models, several competitive, open-source alternatives have emerged. 
Both industry~\cite{touvron2023llama, zhang2022opt, touvron2023llama2} as well as collaborative efforts~\cite{workshop2023bloom} have open-sourced LLMs aiming to improve public access.
However, these alternatives still require significant computing and energy resources to match the state-of-the-art performance.

\textbf{Cost and Energy:} Several works~\cite{zhao2023survey, naveed2023comprehensive} have analyzed the financial costs required to train and operate LLMs without delving deep into their long term ethical implications.
Other works~\cite{faiz2024llmcarbon, strubell2019energy, roy2020green} have looked at the energy costs of these models and the potential environmental impact. Compared to these, we analyze the cost and energy usage from a collective control perspective, asking ethical questions such as, will the future of LLMs be democratic, fostering a society where everyone can use and develop them? If not, which groups of population and organizations determine who has access? and how will this impact the broader society in the long run?

\textbf{LLM Cost Reduction:} Recently, several papers have analyzed methods to improve the efficiency of serving LLMs. 
One such method is quantization which reduces the precision of LLM weights~\cite{Neves_2023}. 
While this reduces the required memory and compute resources needed to serve LLMs~\cite{zhao2023atom}, it also reduces accuracy and precision, leading to concerns about model quality and equity. 
For example, models that use quantization may be less powerful compared to models that do not.
In addition, quantization is only used for LLM serving, whereas training still requires significant computational resources~\cite{dettmers2023qlora}.

% Advancements in efficiency helps us move towards decreasing LLM costs. 
% However, as we will show, we are still far from LLMs that are widely accessible. 
%While we do not present any new methods for LLM efficiency, 
%We analyze models that take advantage of efficiency methods and highlight the gap that still exists in LLM accessibility. 

\textbf{LLMs, Society, and Ethics}: Attempts have been made in prior work to analyse the influence of AI models in shaping society~\cite{bommasani2022opportunities, roy2020green}. 
% LLMs are a powerful technology that can augment human skills to create new opportunities, akin to the development of steam engines and the internet. 
% Inequity in their access can lead to the potential risk of increased concentration of ownership and power to a small group of corporations 
 Several works have noted the dominance of English in Language Models, making them almost exclusive to European and North American Societies~\cite{langLLM1, ramesh2023fairness}. Ethical concerns have been raised over the exposure of toxic online content among underpaid moderators, and the significant impact on their mental health during the data collection process for these LLM pipelines ~\cite{online_toxic}. Moreover, AI technology is believed to be moving towards a "Turing trap" with excessive focus on human-like AI, with an eventual goal of replacing human labour. This process centralises power towards those who control the technology~\cite{brynjolfsson2022turing}. 
 Hence, it is essential to analyse the distribution of LLM access across different socio-economic groups to promote equity in access to this powerful technology. 
To date, there has been limited research in this area~\cite{Bender:FaccT:2021}. 
 Our project aims to fill this gap by drawing relevant insights about the ethics of inequity of this technology.
 %Our project aims to fill this gap by 1. collecting relevant data for socio-economic structures of different sizes like countries, startups, research labs, and individuals, 2. projecting their resource constraints on access to expensive models and, 3. drawing relevant insights about the ethics of inequity.

\section{Methodology}
\label{sec:methods}

In this section, we outline how we identify and analyze the financial requirements needed for training and serving LLMs. We also detail our method of quantifying distribution of financial resources around the world.
% by looking at three types of LLM stakeholders: (1) companies, (2) research institutions, and (3) individuals. We compare the financial costs of training and serving LLMs to the financial resources of these groups to uncover the accessibility of LLMs for different communities.  
We rely entirely on published and open data.

% As we are aggregating this data from other sources, the accuracy of these numbers depends on the publisher.
% We pick the most reputable sources wherever possible.

\begin{table}
\centering
    \caption{Details of LLMs used for our analysis.}
    \def\arraystretch{1.1}
    \begin{tabular}{|c|c|c|c|c|c|}
        \hline
        \multirow{2}{*}{\textbf{Model}} &  \multirow{2}{*}{\textbf{Size (Params)}} &  \multirow{2}{*}{\textbf{Release}}  & \multicolumn{2}{|c|}{\textbf{Cost}} & \textbf{Performance}\\ \cline{4-5}
         &  & & \textbf{Train (\$)} & \textbf{Serve (\$/day)} & \textbf{(MMLU \%)} \\ \hline
          \hline
         T5~\cite{T5_model_paper} &  Small (11B) &  Oct 2019 & 2 Mil & 36 & 25.9 \\ \hline
         LLaMA-2~\cite{zhang2022opt} &  Small (7B) &  July 2023 & 5.52 Mil & 18 & 45.3 \\ \hline
         GPT-3~\cite{brown2020language} &  Medium (175B) &  June 2020 & 5 Mil & 180 & 43.9\\ \hline
         BLOOM~\cite{workshop2023bloom} &  Medium (176B) &  July 2022 & 5 Mil & 180 & 39.1 \\ \hline

         PaLM~\cite{chowdhery2022palm} &  Large (540B)&  April 2022 & 23.1 Mil & 500 & 71.3 \\ \hline
         GPT-4~\cite{openai2023gpt4} & Large (1000B)& March 2023 & 100 Mil & 900 & 86.4 \\ \hline
    \end{tabular}
    \label{tab:llm_costs}
    
\end{table}

\subsection{Models}
We focus on 6 different models spanning three categories: \circled{1} Small ($\sim$10B parameters), \circled{2} Medium ($\sim$100B parameters), and \circled{3} Large LLMs (>500B parameters). We only consider models that we have observed to be commonly used or cited by the general public.  

\noindent\textbf{Cost}: Table~\ref{tab:llm_costs} contains our list of LLMs and their associated costs.
To identify the training and serving costs of these models, we analyze the resource data published by model creators. 
For unknown training costs, we base our analysis on GPU rental prices on the cloud~\cite{cost_nlp_models}.
We use an optimistic estimate of $\$1.5 /GPU/hr$~\cite{gpu_pricing1, gpu_pricing2} for a single $A100$ GPU served on the cloud, and multiply that with the GPU hours required for training the model. For example, LLaMA-2 13B required over 368,000 GPU hours to train~\cite{llama_gpu_hours} making the total training cost $368000 * 1.5 \simeq \$5.52M$.
An alternative approach would be to calculate costs if organizations performed the training in-house.
This would require calculating the costs of purchasing GPUs, constructing a datacenter, and operating the datacenter, which is complicated and error-prone.

We estimate the inference costs by computing the memory required to load the model, the number of A100 GPUs (80GB) required to satisfy the memory constraints~\cite{huggingface_optimize}, and the daily cost of serving these models on the cloud. For example, inference with GPT-3 requires 175 billion 16-bit floating point parameters. The total number of A100s required would be $(175\ billion * (16/8)) / 80 \simeq 5\ GPUs$. The cost of serving this for 1 day would be $5 * \$1.5 * 24 = \$180$. 

\noindent\textbf{Performance}: To compare the performance of these models, we rely on the MMLU~\cite{hendrycks2021measuring} benchmark.
This benchmark consists of multiple choice tests spanning 57 different domains, intending to holistically capture the knowledge and reasoning capabilities of different models.
A score of 25\% implies that a model is as good as random chance.

Table~\ref{tab:llm_costs} contains the publicly reported performance of T5~\cite{chung2022scaling}, GPT-3~\cite{hendrycks2021measuring}, BLOOM~\cite{wu2023bloomberggpt}, LLaMA-2~\cite{zhang2022opt}, PaLM~\cite{chung2022scaling}, and GPT-4~\cite{openai2023gpt4}.
These numbers are gathered from different sources and the exact evaluation setup may not match.
However, this gives us a rough scale to compare the models.

\ignore{
The MMLU score is a scale from 0\% to 100\%, however, it is not a perfectly linear scale.
As scores approach 100\%, the difficulty in developing a model that beats the state-of-the-art becomes harder.
For example, the performance gap between 80\% and 60\% is significantly higher than 60\% and 40\%, despite having the same absolute difference.

To highlight this gap, we define the "Power" of an LLM using the following formula:\\
$$Power = \frac{1}{(1 - Score)}$$ \\
Using this formula, an LLM which performs as well as random chance would have a power of 0, while an LLM which answers perfectly would have an infinite power.}

% \begin{table}
% \centering
%     \caption{Details of LLMs used for our analysis.}
%     \begin{tabular}{c|c|c}
%         \textbf{Model} &  \textbf{Size} &  \textbf{Release Date}  \\ \hline
%         & &\\
%          T5&  small &  Oct 2019 \\
%          LLaMA2&  small&  July 2023 \\
%          GPT-3&  medium&  July 2022 \\
%          BLOOM&  medium&  July 2022 \\
%          PaLM&  large&  April 2022 \\
%          GPT-4&  large&  March 2023 \\
%     \end{tabular}
%     \label{tab:my_label}
    
% \end{table}

\subsection{Populations}
For this project, we choose stakeholders which we divide into three categories:
\begin{enumerate}
    \item \textbf{Businesses}: As LLMs have powerful capabilities in a wide range of tasks, from text translation~\cite{zhu2023multilingual, wang2023document} to mathematical reasoning~\cite{imani2023mathprompter}, many companies will want to adopt LLMs and build businesses around them. Those with the resources to train their own LLMs will hold a significant advantage in integrating AI into their products over companies that do not have such resources.
    \item \textbf{Research Institutions}: The ability for researchers to participate in development of NLP and AI algorithms ensure that LLMs properly serve their communities. But this depends largely on the accessibility of LLMs for their research. If LLMs are not accessible to localised researchers, it is difficult for the overall development of LLM to align with the needs of specific communities. 
    \item \textbf{Individuals}: Many LLM-based products, such as ChatGPT, Gemini, and Bing Chat, are extremely useful in everyday life. They have the power to significantly increase the productivity of skilled workers in domains like writing, journalism and even research.  However, to access the powerful capabilities of LLMs, individuals need to pay usage fees. The price tag that comes with LLM inference plays a significant role in determining which individuals will be end users of LLMs.
\end{enumerate}

For businesses, we omit the larger corporations that already hold significant monopoly and instead evaluate financial resources by comparing the venture capital (VC) funding across different countries. VC funding allows us to capture the financing power of small business that have high growth potential. We extract our VC funding data from~\cite{seed_fund_vc} and~\cite{dealroom_vc}. 
% and filter the seed fund investments during 2022 for all the available countries. Our data source collects venture capital investments for Early and Late-stage startups in Europe and North America. We were unable to find reliable VC funding data sources for African and South American countries. Hence, we stick to the above reliable source which only represents a fraction of the world. We also choose only seed funds when compared to late-stage startups, since we are interested in the earliest footsteps of the startup world.
To examine the financial resources of research institutions, we use national research budgets as a proxy
%Very little comprehensive data is published around funding levels of research institutions. However, many universities and national laboratories receive funding from their country's research grants. These are allocated through national budgets, and since this data is publicly available, we choose to use it to assess research institution funding.
as obtained from the World Bank~\cite{academic_funding}.
% The World Bank publishes the amount that countries invest in research as a percentage of their GDP.
We acknowledge that research budgets may not accurately describe the amount a country will spend on developing LLMs as certain countries may prioritize other research goals, but we begin with these numbers as a starting point for analyzing how likely a country could invest in AI endeavours. 
We finally look at how accessible LLMs would be to individuals of different nations. 
For this, we take the average and median income of people around the world~\cite{world_pop_review}.
Similar to research budgets, this allows us to see how likely the average citizen of the country would have access to this technology.
%\todo{TODO:} Paragraph about comparing against the income of individuals around the world using median income per country.

\section{Results} \label{sec:results}
% \section{Control of LLM technology} \label{sec:results_control}
% We now present our findings on the distribution of financial viability for training and inference of LLMs around the world.

\begin{figure}[t]%
    \centering

    \begin{subfigure}{0.47\linewidth}
    \includegraphics[width=1\linewidth]{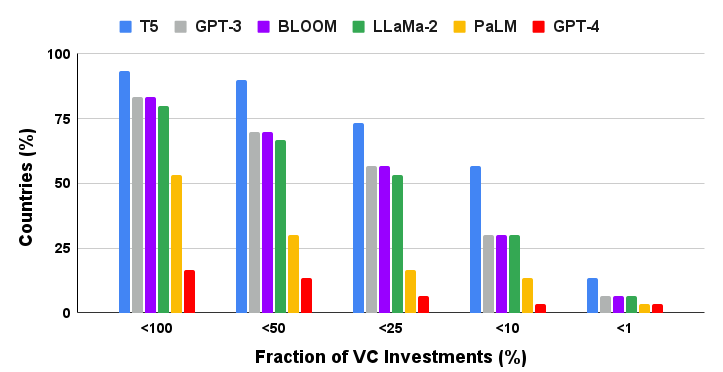}
    \caption{Training costs as a fraction of VC funding.}\label{fig:vc_budget}
    \end{subfigure}
    \qquad
    \begin{subfigure}{0.47\linewidth}
    \hspace*{-0.25in}
    \includegraphics[width=1.2\linewidth]{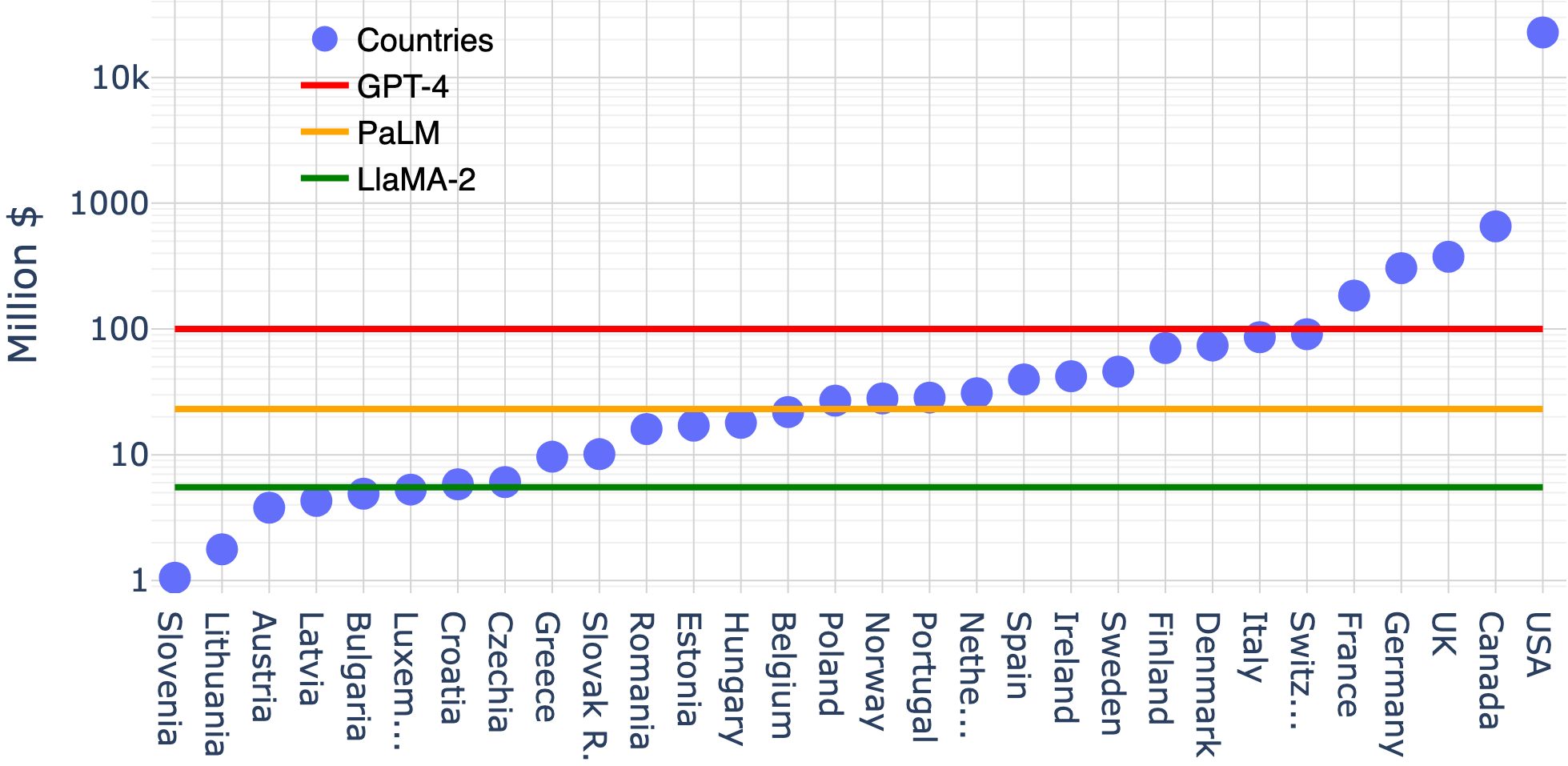}
    \caption{Countries and their VC funding for 2022.}\label{fig:vc_comparison}
    \end{subfigure}

    \begin{subfigure}{0.47\linewidth}
    \includegraphics[width=\linewidth]{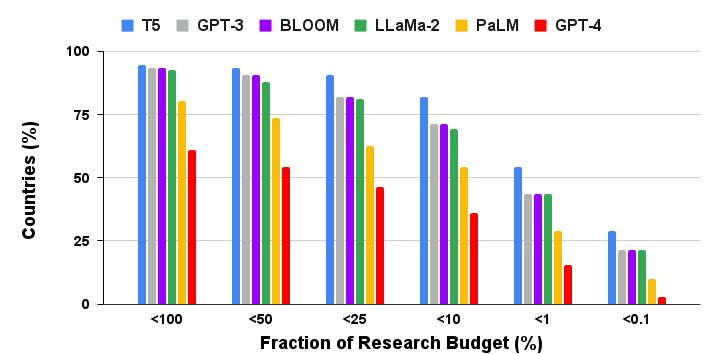}
    \caption{Training costs as a fraction of national research budgets.}\label{fig:research_budget}
    \end{subfigure}
    \qquad
    \begin{subfigure}{0.47\linewidth}
    \includegraphics[width=\linewidth]{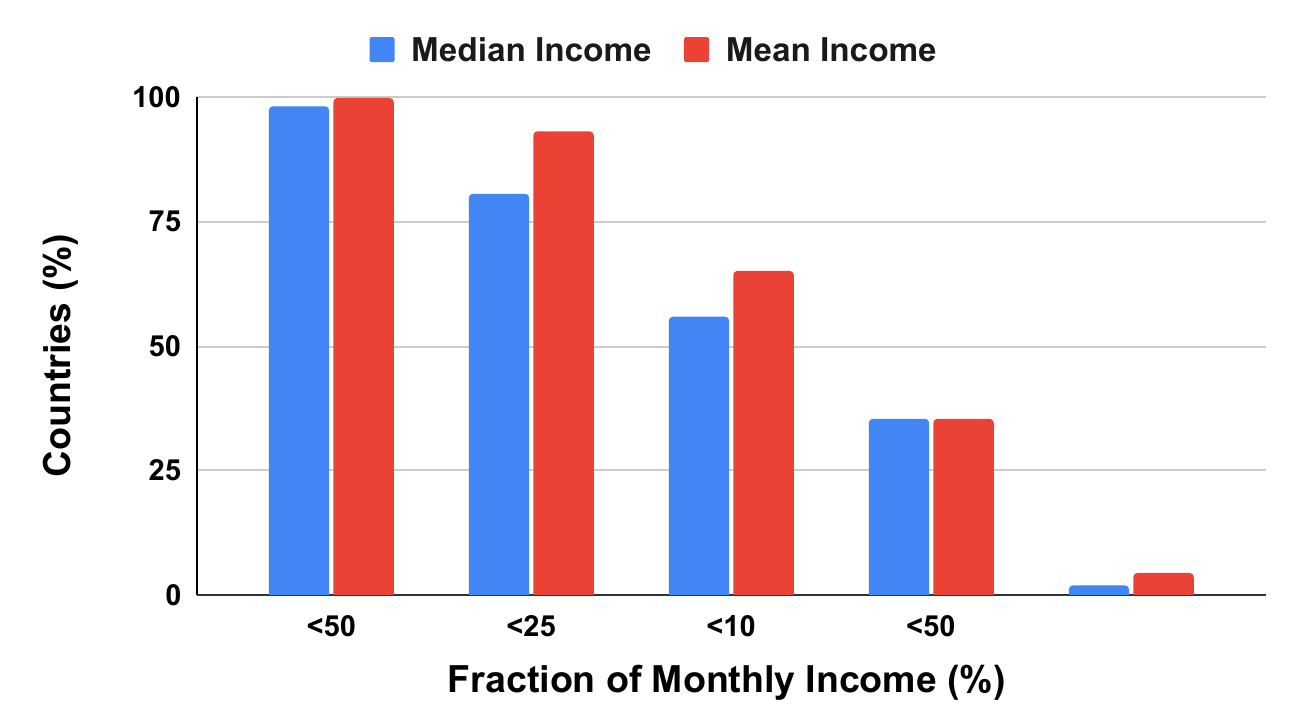}
    \caption{LLM subscription costs as a fraction of mean/median income.}\label{fig:individual}
    \end{subfigure}

    \caption{Model training and subscription costs  for startups, research institutions.}%
    \label{fig:results}
\end{figure}

\subsection{Businesses}

Figures \ref{fig:vc_budget} and \ref{fig:vc_comparison} show VC funding budget compared to the costs of training small, medium, and large LLMs. These figures focuses on early and late-stage startups in Europe and North America in 2022. From Figure \ref{fig:vc_budget}, we notice that United States and Canada are the only nations that could train GPT-4 with less than 25\% of their total seed fund investments. In Figure \ref{fig:vc_comparison}, we use a log scale to represent the investments on the Y-axis due to their highly skewed nature. The United States receives over 100 times more VC funding than France or Italy and is particularly capable of training the largest LLMs.

\subsection{Research Institutions}

We analyzed 149 countries whose national research budgets were available.
To avoid the impact of temporary economic downturns, we picked the largest amount that each country spent on research within the span of 1996 - 2021. Figure~\ref{fig:research_budget} shows the fraction of the research budgets that these countries would have to spend to be able to afford training the various models.
58 countries would not be able to train GPT-4 even if they spent their entire research budget.
Practically, countries will have various different research priorities, like healthcare, agriculture, etc., which will divert certain amounts of their budget.
Only 4 countries were able to afford training GPT-4 using less that 0.1\% of their research budget, namely USA, China, Japan, and Germany.

\subsection{Individuals} \label{sec:indivudals} 
Based on the subscription fee for ChatGPT Plus~\cite{openaiPricing}, we estimate the cost of using LLM services to be \$20/month. We compare this pricing to the median and mean income of countries listed in~\cite{world_pop_review}. If individuals are willing to place their entire monthly salary into buying LLM services, people from all countries (present in~\cite{world_pop_review}) can afford LLM services on an average. However, it is more realistic to estimate that individuals are willing to put around 10\% of their monthly income into subscription services according to recent analysis ~\cite{sub_crr}. At this rate, only individuals living in the top 56\% of countries can afford LLM services.

\section{Discussion} \label{sec:discussion}

\begin{figure}%
    \centering

    \begin{subfigure}{0.38\linewidth}
    \includegraphics[width=\linewidth]{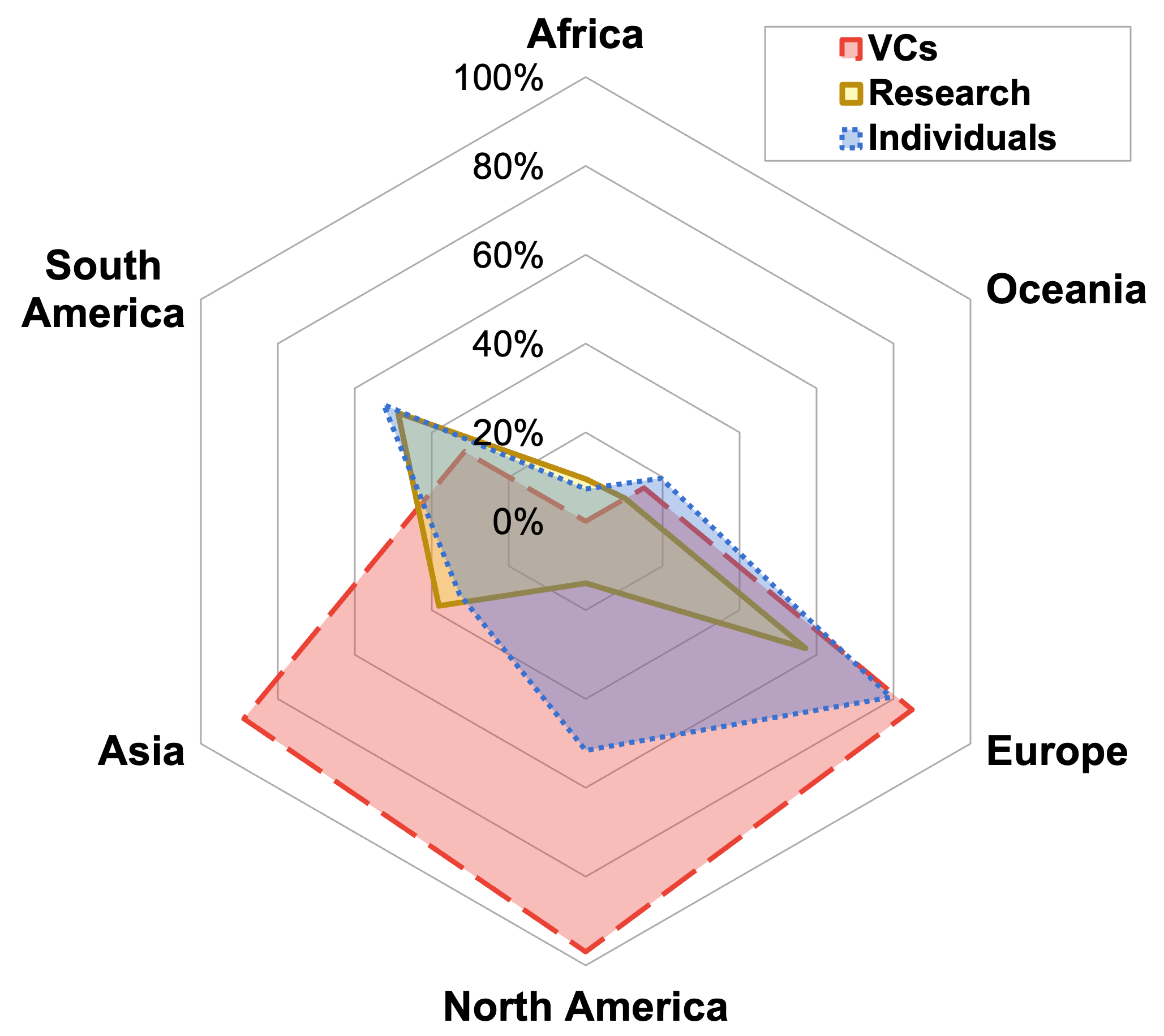}
    \caption{LLM training and inference viability across continents.}\label{fig:radar}
    \end{subfigure}
    \hfill%
    \begin{subfigure}{0.58\linewidth}
    \includegraphics[width=1\linewidth]{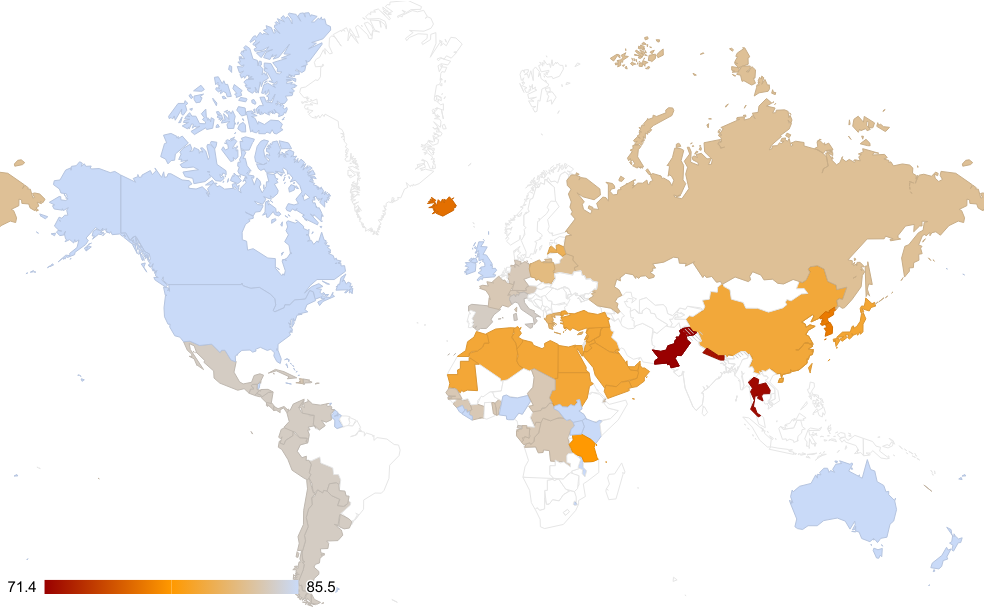}
    \caption{MMLU performance of GPT-4 by language. Each country is shaded by its MMLU score for its national language.}\label{fig:mmlu}
    \end{subfigure}
    \qquad

    \caption{Distribution of model training and subscription costs and model language capabilities around the world.}%
    \label{fig:discussion}
\end{figure}

\subsection{Who holds control of LLMs?}

From our data in Section~\ref{sec:results}, we see that the majority of financial viability for both training and operating LLMs are concentrated in wealthy nations, such as the United States. Figure~\ref{fig:radar} combines data from Figure~\ref{fig:results} by showing the average amount remaining if countries spend up to a generous 50\% of their VC funding and research budgets on training models and if individuals spend up to 10\% of their income on LLM subscriptions. The figure highlights the disparity of financial viability between different countries and presents the concentration of VC, research, and individual financial viability in Asian, North American, and European countries. This inequality is also reflected in the distribution of Massive Multitask Language Understand (MMLU) performance around the world. The distribution of MMLU performance for GPT-4 in Figure~\ref{fig:mmlu} shows the majority of LLM language power lies in English-speaking countries~\cite{openai2023gpt4}.

\subsection{Expanding LLMs to All}
\paragraph{Why concentration of LLM power is concerning:}

The high compute costs of LLMs has led to a concentration of LLM control among the most wealthy countries. This raises the question: In what ways have the strong biases in access and control of these models impacted society? Upon analysis, we find that despite efforts from several wealthy corporations to produce LLMs that are equitable and fair, these statistical models have been found to produce and amplify social biases ~\cite{gallegos2023bias}. Many LLMs hold cultural values and political biases reflecting those of model trainers ~\cite{liu2022quantifying}. The occurrence of hate speech, misinformation, and gender or cultural biases can differ in LLMs depending on how training data is collected and how the model is trained~\cite{feng2023pretraining, durmus2023measuring}.  If LLM training is constrained to a few most highly-resourced countries, LLMs may misrepresent minorities and show biases against those from less affluent countries.

Another important observation to consider in the discussion around equity is that the top LLM-based services now are offered through payment plans. As highlighted in Figure~\ref{fig:individual}, these pay-walled services are not accessible to all individuals. Pay-for-use LLMs may further increase the economic divide in the world, especially given the high potential for LLMs to increase productivity for users who have access~\cite{meyer2023chatgpt}. If only the rich are able to access LLMs and improve their productivity, will we exacerbate the current inequality in the world economy?

\paragraph{Biases in language:}

Figure ~\ref{fig:mmlu} shows the MMLU performance of GPT-4, the latest and the most powerful language model across different languages. MMLU is the measure of the model's multitask accuracy across 57 tasks ~\cite{hendrycks2021measuring}.  Blue shows the regions (and their official language) with the highest MMLU scores. It is clear from the figure that English dominates as the language that performs best at multiple language tasks. The disparity in language performance brings us to several concerns. What are the implications of LLM impact on other highly used languages like Urdu, Thai, Hindi, etc.? Would people using these languages not be able to make effective use of the technology? What would the future look like when most languages are left behind by such a powerful technology?

\paragraph{Expanding LLM accessibility:}
Given the hurdle that LLM compute costs create for accessing LLMs and the ethical concerns they bring, here we offer a few ideas on how we might increase accessibility of LLMs. 

\emph{1. Invest in developing methods to decrease the compute costs of LLMs.} \\
Some research has explored how to improve LLM training and serving efficiency, such as model quantization~\cite{Neves_2023}, distillation~\cite{gou2021knowledge}, and pruning~\cite{ma2024llm}. We encourage the community to continue investing in such methods and identifying ways to further decrease LLM compute costs and to make AI more green~\cite{schwartz2020green}. It is crucial for us to consider how we can increase accessibility for LLMs today before expanding the models and making them more resource intensive and more difficult for others to train and serve.

\emph{2. Encourage cross-cultural and international collaborations in development of LLMs.} \\
Much of LLM training today is concentrated within select companies within certain countries (Figures~\ref{fig:results},~\ref{fig:discussion}). We can lessen this by incorporating cross-cultural co-design and collaboration. Cross-cultural development and research can help create models with an increased understanding of the different cultures of the end users. Increasing diversity in the LLM research ecosystem may also help researchers gain a deeper understanding of compute constraints within low-resourced communities and the potential impacts of decisions being made during LLM training.
    
\emph{3. Advocate for policies and initiatives that increase fair access to LLMs} \\
Ultimately, changes in who holds power over LLMs will require policies and intiatives that enforce fair practice and deployment of LLMs. We call the community to look for ways to enact policies that will allow for increased LLM access, such as ensuring LLM companies develop fair payment plans that consider economic viability globally and not just among the wealthy.

\subsection{Limitations}
% Given the short time frame of our class,
We chose to rely on existing datasets to glean financial data as outline in Section~\ref{sec:methods}, and our results only capture groups and countries covered in those datasets. For individual income and academic research institutions, we take aggregates. We acknowledge that the choice of categories and aggregation has implications and that there are a wide variance of economic conditions within each group. We encourage further research to better capture the nuances within each group and their implications on LLM accessibility. With the introductory analysis and awareness on LLM access discussed in this article, we hope to provide an inspiring starting point for AI startups and researchers, to address these issues more concretely.

% \subsection{Positionality Statement}

\section{Conclusion} \label{sec:conclusion}

We have found that access to training and inference for LLMs is highly concentrated within the hands of a few countries. With the exorbitant costs of training LLMs, companies, research institutions, and individuals with the highest economic earnings represent the majority of those with abilities to train LLMs. Similarly, costs for inference constrain using LLMs to those in the top earning countries. This leads to a concerning concentration of LLM power among a small group. We discuss several implications of the observed inequity and pose important questions to consider looking forward. We propose initial ideas for how to combat this monopolization. We also encourage the research community to continue examining ways to decrease the costs of training LLMs and to allow those in lower-resourced countries easier access to LLM inference. 

%\bibliography{cse581}
\printbibliography

\end{document}